# Plasma Panel Detectors for MIP Detection for the SLHC and a Test Chamber Design

R. Ball[1], J. W. Chapman[1], E. Etzion[3], P.S. Friedman[2], D. S. Levin[1], M. Ben Moshe[3], C. Weaverdyck[1], B. Zhou[1]

[1]Department of Physics, University of Michigan Ann Arbor, MI
[2]Integrated Sensors, LLC, Toledo, OH
[3]Raymond and Beverly Sackler School of Physics and Astronomy, Tel Aviv University, Israel

*Abstract*–**Performance demands for high and super-high luminosity at the LHC ( up to $10^{35}$ cm$^{-2}$ sec$^{-1}$ after the 2017 shutdown) and at future colliders demand high resolution tracking detectors with very fast time response and excellent temporal and spatial resolution. We are investigating a new radiation detector technology based on Plasma Display Panels (PDP), the underlying engine of panel plasma television displays. The design and production of PDPs is supported by four decades of industrial development. Emerging from this television technology is the Plasma Panel Sensor (PPS), a novel variant of the micropattern radiation detector. The PPS is fundamentally an array of micro-Geiger plasma discharge cells operating in a non-ageing, hermetically sealed gas mixture. We report on the PPS development program, including design of a PPS Test Cell.**

## INTRODUCTION

The demanding detector performance requirements for high (> 3 ×$10^{34}$ cm$^{-2}$ sec$^{-1}$) and super-high ($10^{35}$ cm$^{-2}$ sec$^{-1}$) luminosities at the LHC and future facilities such as the International Linear Collider have motivated our investigations into a radiation detector technology based on Plasma Display Panels (PDP). The PDP's are the principal component of flat panel plasma television displays. Their design and production is supported by an extensive industrial infrastructure with four decades of development. As display units, plasma panels have proven reliability, durability, very long lifetimes coupled with low costs with applications in both commercial and military sectors. Our objective is to develop from the PDP technology, a novel micro-pattern radiation detector – herein referred to as Plasma Panel Sensors (PPS) (references 1-3). The PPS is intended to utilize and benefit from many of the key attributes of plasma panels.

A PDP comprises millions of cells per square meter, each of which can, when provided with a signal pulse, initiate and sustain a plasma discharge. As a plasma panel detector, a PDP cell can be biased to discharge when a free-electron is generated or injected into the gas. Therefore the PPS, as a *redesigned PDP*, functions as a highly integrated array of *parallel* pixel-sensor-elements or cells, each independently capable of detecting *single* free-electrons generated within the cell by incident ionizing radiation. While the emphasis of this paper is on muon (minimum ionizing particle) radiation detection, the PPS with an appropriate front end might also be configured as photodetector.

Qualitatively, the potential *performance* attributes of the PPS are fast response (order $10^{-11}$ s), high gain, high data rate capability (> $10^9$ Hz/cm$^2$), extremely low power consumption, combined with an RMS spatial resolution of well under 100 microns. Mechanically, a PPS detector is comprised of radiation-hard, inert components, hermetically sealed and notably having no external gas flow-supply system. It is lightweight and structurally robust.

Costs of PDP television production are very well established and have been dropping year by year. Current retail market value of PDPs (4) with electronics is at \$0.30 inch$^{-2}$. While it is impossible to predict the much lower volume production costs for PPS detectors, their design, components and manufacture are intended to have maximal overlap with PDPs. They can incorporate the same types of PDP glass, electrode materials and industrial photolithography processing, similar gas mixtures and panel fabrication and sealing processes. The electronics and power supply requirements of PPS detectors are not expected exceed those of commercial PDPs, and in many respects will be simpler.

PPS attributes will be considered below in greater detail. A final section describes the design of a PPS test cell.

## I. PLASMA PANEL SENSORS DETECTOR DESCRIPTION

### A. Plasma Display Panel televisions

The PPS is based on the Plasma Panel Display (PDP). The basic element of PDPs consists of orthogonal arrays of electrodes deposited (or etched) onto glass substrates, separated by a gas discharge gap. Figure 1 illustrates the basic features of an AC-type, PDP assembly. The pixel or cell contains a phosphor coated wall structure enclosing a plasma discharge cell. A PDP is comprised of millions of cells per square meter, each of which can initiate and sustain a plasma discharge.

The cells are quite small: on the order of 200 μm in each dimension for HDTV, although PDPs with cell dimensions on the order of 100 μm have been made for military applications. Because of the small electrode gaps, large electric fields arise with only a few hundred volts of bias. At least three cells, each associated to a specific color (e.g. RGB), comprise a pixel. The plasma discharges are usually made in mixtures of noble gasses: typically Xe and Ne gas at about 500 torr. It is important to note that the gas in a PDP is permanently and hermetically sealed in the panel's glass envelope. The discharge produces VUV photons that excite phosphors in the cells and produces the bright colors characteristic of plasma TV. At any instant in a displayed image many of the pixels have at least one cell on, so the PDP electronics must individually address, refresh and sustain these discharges, while quickly suppressing cells that must change state by *"erasing" their stored charge on the surface dielectric*.

For plasma panel detectors, a PDP cell can be biased to discharge when a free-electron is generated or injected into the gas. The PPS, as a *redesigned PDP*, functions as a highly integrated array of *parallel* pixel-sensor-elements or cells, each independently capable of detecting *single* free-electrons generated within the cell by incident ionizing radiation. Such electrons then undergo rapid electron multiplication resulting in an avalanche and discharge that can be confined to the local pixel cell space. For all PDP products, this process is self-limiting and self-contained by various means, one of the most important being a localized impedance at each cell. The total charge available to produce a signal is that stored by the cell's internal capacitance and determines an effective gain. This gain therefore depends on details of cell geometry and materials, and are estimated (see below) for a PPS with a 100 μm pixel pitch to be on the order of $10^6$. Since the cell is operated above the proportional mode, in essence it may be viewed as a micro-Geiger counter. The signal pulse will be independent of the number of initiating free electrons, rendering therefore the PPS as *intrinsically digital.* The gain may be sufficient to obviate signal amplification electronics.

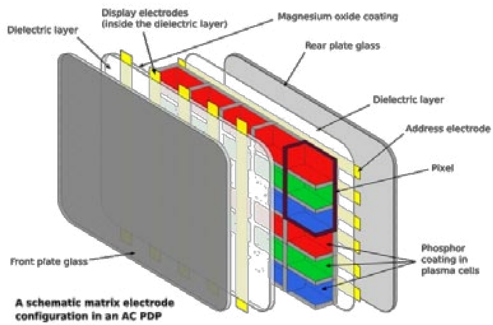

**Figure 1**: Structure of an AC type Plasma Display Panel showing the basic elements. The columnar plasma discharge region is defined by the crossing electrodes. The gas mixture comprised of noble gasses is hermetically sealed in the glass envelope. In a PPS device phosphors, MgO coatings and dielectric layers over the display electrode are absent.

*PPS cell geometry*

The cell geometry includes the dimensions of the electrodes, their pitch and vertical spacing, and dimensions of cell walls. For MIP detection the electrode layout should have a "vertical" drift region on the order of 2 to 3 mm and a transverse electric field avalanche region of 50-100 μm. The drift region is required to ensure sufficient probability that a passing particle will produce at least one ion-pair. In Ar at 1 atm, for example, a MIP produces about 25- 30 interaction "clusters" per cm yielding at least one ion-pair. On average, 100 primary and secondary ion-pairs per cm are produced (5). The cluster generation is Poisson distributed, thus in 2 mm and 3 mm drift regions at 500 torr (the anticipated PPS pressure) the probability to generate at least one cluster is ~ 99%. Due to the large gain, a single ion-pair can initiate a signal, suggesting that a drift region of ~ 3 mm is sufficient to produce signals with very high efficiency.

The PPS electrode geometries are depicted in a number of ways. A 2D view of a simplified cell with one readout coordinate is shown in Figure 2. This electrode geometry evokes that of a micro-strip gas counter. Here, however, the electrodes are strips of limited depth of order hundreds of microns into the page. The purpose of this representation is to establish the electric field lines and electron drift trajectories. These have been computed using the Garfield (6) and Maxwell-2D (7) programs (Figure 3).

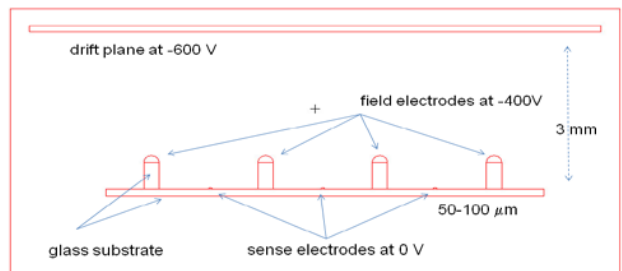

**Figure 2:** 2D side view of an initial conception of electrode geometry along one coordinate of a PPS cell. In this view the electrode layout evokes a Microstrip gas counter. The electrodes however are not long strips, but have a limited depth into the page of order hundred microns. Specified potentials are suggestive.

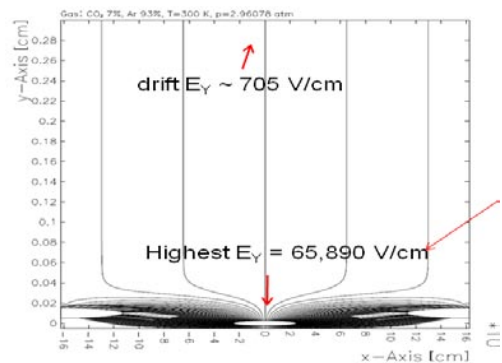

**Figure 3:** The electric fields in the drift and avalanche directions for the round electrodes of the cell represented in Figure 2 (x-axis units in cm*$10^{-3}$). The field between the avalanche and sense electrodes reaches a similar value as does the field from the drift region as it nears the sense electrode.

A *conceptual* representation of a cell configuration is shown in Figure 4. In this depiction the cell is defined by a local electrode arrangement with an intrinsic capacitance and an embedded resistance in the high voltage feeds (the X-electrode in this figure) to each cell. That is the pixel here is represented by the proximity of the two electrodes defining a capacitor, the high voltage (or discharge) side of each being fed by a resistance. The resistance drops the high voltage at discharge and effectively terminates and localizes the discharge. This is one of the operating principals of resistive plate counters (RPCs). The effectiveness of this resistance is investigated with SPICE (8) simulations in a section below.

In a practical PPS section, the embedded resistances are not expected to be distributed on the same plane as the pixels. They rather will be formed by resistive depositions that lie below the discharge electrode. An illustration of this layered structure (Figure 5) depicts an electrode configuration where embedded cell resistors derive from a resistive layer bridging between the discharge electrode and the high voltage line. The discharge electrode assembly is shown recessed into the substrate. In an alternative design this discharge electrode and resistive layer are located on the substrate surface, leaving the electrode slightly elevated with respect to the sense line. Simulations indicate that this elevation has significant impact on the drift field. While the dimensions shown are illustrative only, electrodes of the sizes indicated (25-75 μm) are easily within current low-cost manufacturing capability.

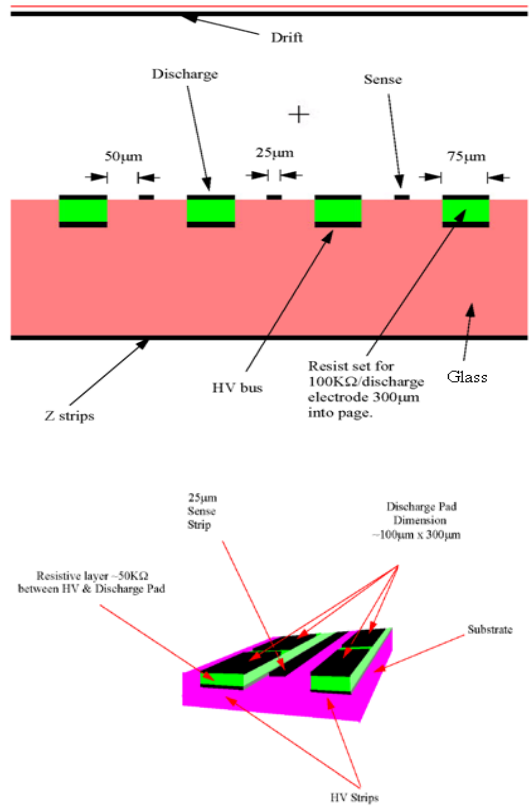

**Figure 5**: (Top) Side view of electrode configuration where embedded cell resistors derive from a resistive layer bridging between the discharge electrode and the high voltage line. The discharge electrodes are shown recessed into the substrate. (Bottom) A perspective view.

## II. PPS MATERIALS

Material selection for PPS modules is driven by the requirement to minimize detector ageing effects over the extensive, $10^5$ hour lifetimes characteristic of plasma displays. Adoption of well established PDP materials and fabrication technique informs the choices for PPSs. The materials requirements are determined by two objectives: radiation hardness and non-ageing, and localization of the discharge. All materials intended for use in PPS are intrinsically non-degrading with exposure to UV and VUV photons and ionizing radiation. PPS devices, like PDPs, incorporate inert, non-reactive and non-polymeric components. In the following subsections, a few specific considerations of material selection for PPS components are considered.

### A. Substrate

The substrate is comprised of plasma display panel glass. An example of one type of glass is Corning 0211 Microsheet Display Glass, commercially available in standard sizes appropriate for the PPS in thicknesses of 50 - 500 μm. This glass is characterized by very low gas permeability - a critical attribute for long duration gas purity.

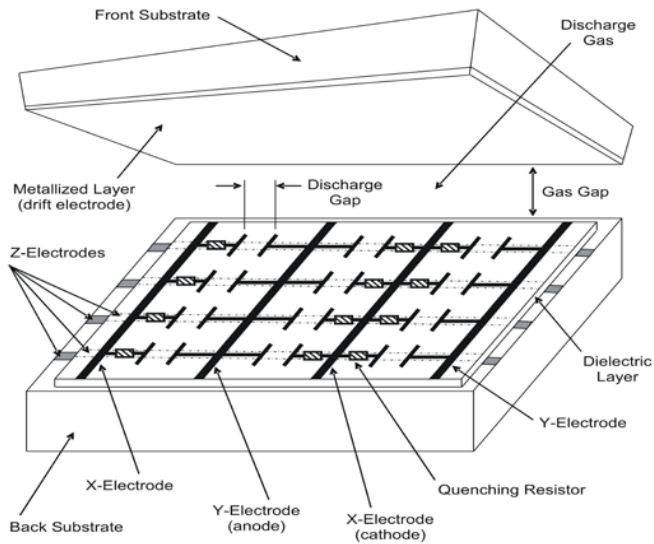

**Figure 4:** Conceptualization, not to scale, of electrode configuration of a photosensitive PPS cell. This is a 4-electrode configuration. The top plate serves as the drift field electrode. The X and Y lines define cells with embedded resistors and avalanche discharge gap regions.

This glass is manufactured in extremely thin sheets (specific gravity=2.53). A muon tracker detector for example featuring multiple sensitive layers can still benefit from a low mass profile. An 8 layer flat panel detector (similar to the number of tracking layers currently used in the ATLAS Muon Spectrometer) fabricated with 100 μm thickness introduces negligible multiple Coulomb scattering to high momentum ( > 10 GeV) muons. In thicknesses of 400 to 500 μm, a variety of glass substrates are available in widths up to 9 meters.

### B. Electrodes and embedded resistors

Electrodes can be fabricated from a variety of durable metals such as molybdenum and tungsten. Durable metals are sputter resistant and less prone to degradation from ion sputtering on the cathodes. Secondly these metals are favored for their lower absorption of UV and VUV photons thus inhibiting photoelectron emission.

Electrode uniformity is necessary to maintain a uniform electric field across a large pixel array. Electrode linewidths of 20 microns or more can be easily manufactured at low cost even in small quantities. Thin film metallization processes on glass or ceramic substrates with ion beam milling can have linewidth uniformities within about ± 2 microns. PPS electrostatic modeling and performance simulations are being done with sense electrodes of 25-30 microns, with gap sizes on the order of 50 to 100 microns. The resulting variation in the electric field across the discharge gap is anticipated to be at most a few percent.

Embedded resistors are fabricated by thick film printing from mixtures of metal oxide particles in melted glass which are laid down on substrates and laser trimmed to form an electrical path with well-controlled conductivity (typically with resistances within 1%).

### C. PPS Gas

Plasma display panels are hermetically-sealed, gas-filled devices, and once installed have demonstrated remarkably long lifetimes. AC-PDP's sold in the 1970's and operating continuously are still functioning today, 30 years later. This leads to an important attribute of PPS detectors: their gas remains sealed inside the envelope and therefore external gas systems are not required for their operation. An *anticipated* problem is that of excited state species (e.g. photons, ions, free electrons and metastables) generated in the gas discharge causing secondary discharges of time-delayed new avalanches. Such secondary discharges could occur at either the original gas discharge pixel site or at neighboring sites.

Therefore, for a PPS to function as particle detector, the gas system must contain a quenching agent and inhibit internal sources of free-electrons, while maintaining sensitivity to electrons created by ionizing radiation passing through the drift region. The PPS gas mixture, like that of PDPs is expected to be comprised of a mono-atomic noble host gas with a quenching agent. The quencher might be another noble gas, or a reasonably stable molecular gas. The choice of gas mixture may also be dictated in part by the requirement to minimize internal sources of free electrons such as might collect on surfaces. A method previously demonstrated (9) to inhibit unwanted free-electrons will be used here to minimize the number of gas-phase metastables, reduce lifetimes of gaseous excited state species and the propagation of VUV emitted photons. This is accomplished via the use of gas-phase quenching and VUV absorbing molecules in a Penning mixture.

Penning mixtures can consist noble host gas (eg., Ar, Ne, Kr, Xe, He ) with a small fraction of gas having a lower ionization potential to depopulate excited metastable states. For example, in monochrome PDP's a common Penning mixture is 0.1% Xe (or Ar) in 99.9% Ne. This mixture reduces the operating voltage while maintaining high amplification and hence good avalanche initiation. Possible Penning dopants (to be used with other host gases) could include: Xe, Kr, Ar, Ne, $CO_2$, $N_2$, $CF_4$, and even Hg. Hg has been used for decades in a variety of plasma discharge devices, including DC-PDP's as well as Hg-Xe arc lamps, and ubiquitously in the common fluorescent light fixture. It is noteworthy that Hg is conductive and could mitigate the problem of stored charge accumulating on the dielectric substrate.

### D. Dielectric supports and barriers

A plasma panel may require internal dielectric structures. Such structures can serve two purposes. The first would be to maintain structural integrity and drift region uniformity. A panel envelope may be operated at an internal pressure of 500-700 torr. Absent internal supports the resulting pressure induced force would deform the panel. Secondly, it may prove necessary to construct a grid of VUV- UV opaque optical barriers surrounding each cell. These barriers are intended to inhibit propagation of VUV and UV photons produced in the electron avalanche to remote regions of the panel, which could, by photoelectric ejection from metal electrodes or direct ionization of the gas cause spurious discharges. That said, a number of methods to obviate the need for such optical barriers, namely a proper choice of gas quenching or VUV absorbing agent and electrode materials, as well as the design of fast quenching electronics are being investigated. It is noted that virtually 100% of all PDP televisions employ an internal barrier structure. However the primary purpose of the barrier in a PDP is to preserve the pixel color saturation and thereby achieve the widest possible color gamut; in other words to prevent UV photons in one cell from also stimulating a different color phosphor in a neighboring cell.

### E. Charge build-up

An anticipated problem associated with dielectric surfaces in a PPS is that of charge build-up or wall charge. All PDP's are designed to maximize wall charge, in part to enhance priming (i.e. free-electron generation), exactly the opposite of what is desired for PPS devices. Since wall charge is the accumulated charge stored on the dielectric surface at the panel gas interface, it is dynamic and therefore can be an internal source of free-electrons. To minimize wall charge, PPS devices ideally ought to minimize 3-dimensional *dielectric* surface structures, such as insulating walls, optical barriers, dielectric spacers, and porous materials. Wall charge that does accumulate needs to be removed almost as quickly as it forms. Three strategies will be investigated: (1) add to the discharge gas a small amount of "conductive" species (e.g. Hg, $NH_3$, etc.) or a positive electron affinity atomic species such as C, F or O via $CO_2$, $O_2$, $CF_4$; (2) carefully select and control dielectric surfaces in contact with the gas for resistivity to allow "bleeding off" of residual charge; and (3) introduce a rapid, periodic, global "charge erase" waveform similar to that used in a PDP-TV to erase the stored charge on each cell so as to turn "off" a lit pixel or erase an entire picture frame.

III. SIGNAL: GAIN, POWER DISSIPATION, TEMPORAL RESPONSE

One interesting property of PDPs is that panel light output and power consumption (dissipation) per unit area is more or less independent of pixel resolution, provided that the other panel material properties are unchanged. Though the pixel pitch might decrease by an order of magnitude or more – e.g. from 1 mm to 0.1 mm, corresponding to the pixel density going from $10^2$ to $10^4$ pixels per $cm^2$ – the light output and power consumption per unit area remains about the same.

To understand the gain, and associated power dissipation we first heuristically estimate the currents and power consumption based on specific assumptions of pixel geometry. This estimate is based on the underlying physics of a discharge event. This estimate is bolstered by a more detailed simulation using Maxwell-2D and SPICE. The former establishes the electric fields and pixel capacitances. The latter is used to model a single cell discharge occurring in a chain of pixel cells of the type to be found in a PPS.

*A. Back-of-Envelope calculation*

The pixel light intensity in a PDP TV-set and the power consumption in a PPS radiation detector are a direct function of the number of charge carriers created in the gas discharge avalanche, which in turn determines the pixel discharge current. Since it is well established experimentally that the power consumption of a PDP is essentially invariant with pixel density, then as the pixel density increases, the power consumed per pixel must decrease by an equivalent amount. Thus for a 100-fold *increase* in pixel density, there must be a corresponding 100-fold *decrease* in power consumption per pixel.

This relationship is determined by the cell capacitance, which can be modeled as a parallel plate capacitor with a specific dielectric constant (k) and dielectric thickness (t). The pixel capacitance is defined as the capacitance of the discharge cathode and sense electrode: $C = k \varepsilon A / t$, where $\varepsilon$ is the permittivity of free space, $8.9 \times 10^{-12}$ F/m, and k is the relative dielectric constant. Additional stray capacitances are ignored here.

Consider a 100 x 100 x 100 µm pixel filled with a gas dielectric. All gases of interest have a dielectric constant very near unity. The substrate glass has higher dielectric constants of approximately 5-6. These parameters set a range of cell capacitances from 1-6 fF. If we select an intermediate value of 3 fF, the bias voltage in a cell, required to turn a pixel "on" and discharge the capacitor, is approximately 300 V, resulting in fields of order $MV/m^2$. Therefore the stored charge in the cell (CV) would be $9 \times 10^{-13}$ coulombs. The maximum amount of available charge for the gas avalanche is set by this geometrically determined stored charge. In a discharge event, approximately one-half of this charge is assumed to flow out of the cell, at which point the electric field is below the threshold necessary to sustain the avalanche at the sense electrode. The reduction in potential across the cell derives from the discharge electrode "pull-up" resistor of tens of thousands of ohms. This resistor, combined with the cell capacitance has at least an order of magnitude longer time constant than the discharge time. The effective gain in this example is set by the amount of released charge: $4 \times 10^{-13}$ Coulomb, or a gain of $\sim 10^6$.

It is important to note that this released charge is independent of the number of free electrons initiating the discharge event; the resultant signal is not proportional to the length of the MIP track through the cell's drift gas volume.

The following estimate of power consumption per unit area includes a design objective where the cell surface density is $10^4$ pixels per $cm^2$, and where each pixel contains an embedded bleed resistance of 50 kΩ. The discharge pulse has two temporal components: a rapid rise time characteristic of avalanche formation of order ps, and a fall or "recovery" time during which the cell recharges through the pull-up resistance. This recovery time is a slowly descending exponential function associated with depleting the ionic space-charge polarization region in the gas. Before a subsequent discharge cycle can take place for a given pixel, the gas space-charge polarization from the previous cycle must be fully dissipated. The time scale for this process is largely determined by the cell's RC-time constant (τ). In this example τ ~ 150 ps. As we are interested in full recovery (and not just 1/e return to full cell re-charge) the recovery time is taken to be more than an order of magnitude higher than the time constant, or ~ 2 ns. (In principal this recovery time corresponds to maximum firing frequency of 500 MHz.) The power dissipation may be considered to be equal to the energy released by a discharge event over this 2 ns interval. This energy is about ½ the stored energy in the pixel capacitance: ½ $CV^2$, assuming that the discharge ends when the voltage drops by half. In this case the released energy is ~ 40 pJ/event. A projected maximum hit rate that might occur in the high radiation Super-LHC environment (10) is 5 KHz/$cm^2$. Taking a factor of four safety (i.e. 20 KHz), this rate yields a power dissipation of ~ 1 µW/$cm^2$.

## IV. MODELING WITH SPICE AND MAXWELL-2D

The electrode configurations and associated signal have been modeled using the Maxwell-2D(7) and SPICE(8) program. The former calculates the electric fields, equipotential surfaces and electrode capacitances. Configurations similar to that shown in Figure 5 were described - including the material dielectric constant of the substrate and electrode shape. The sense electrode capacitance was determined to be approximately ~ 2.5 fF, assuming a glass dielectric of permittivity 5.5. This value is well within the range of the back of the envelope estimation described above. The SPICE modeling and analysis package was then invoked to compute the temporal profile of the high voltage lines and signals. A schematic diagram (Figure 6) describes one chain of 13 cells, including the capacitances of the cell electrodes, orthogonal Z-strip electrodes and various stray capacitances, and all discharge line resistances. The intrinsic cell capacitance was taken to be 3 fF.

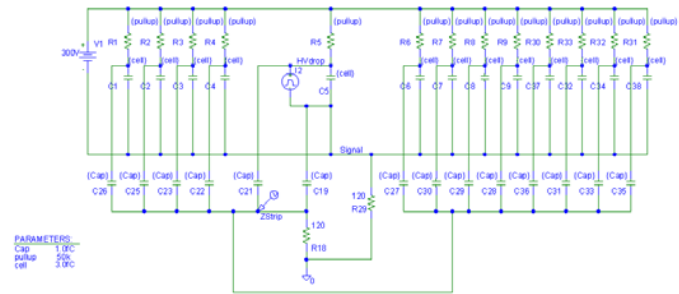

**Figure 6:** Schematic used in SPICE simulation of a 13 cell chain. Resistances labeled "pull-up" correspond to the embedded discharge resistance in Figure 5. The discharge of a "hit" cell is represented by a fast (few ps) current pulse (I2) and set to ½ the stored charge. The cell capacitance is set to 3 fF and the discharge resistance to 50KΩ.

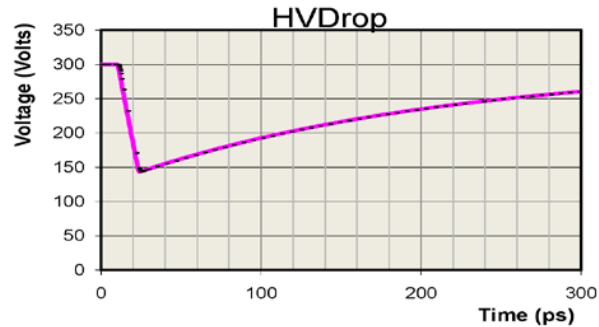

**Figure 7:** The time profile of the high voltage drop across the hit cell in the multi-cell chain in Figure 6. The rise and fall times reflect the cell capacitances and resistances. The drop to ½ the cell potential occurs with a (10%-90%) rise time of ~ 8 ps. The fall time (1/e return to baseline) is ~250 ps. The high voltage *across adjacent cells* remains unchanged at 300 V, indicating that the discharge remains localized.

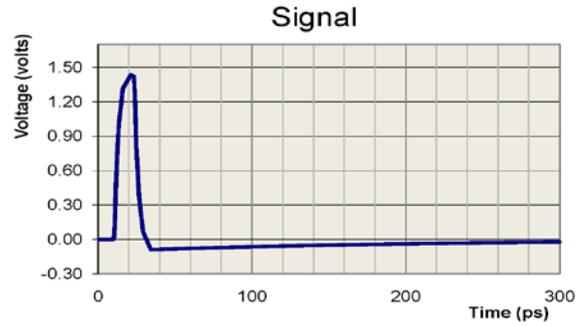

**Figure 8:** The time profile of the sense line signal produced by the hit cell in the multi-cell chain in Figure 6. The signal appears across the 120 Ω output impedance.

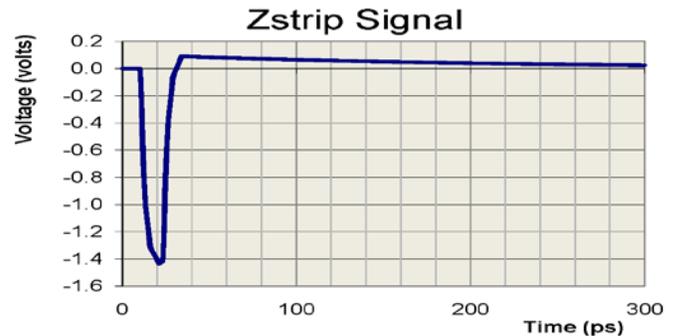

**Figure 9:** The time profile of the pulse on an orthogonal Z-strip produced by the hit cell above.

### A. SPICE simulation results

The SPICE simulation has the following assumptions: The voltage needed to extinguish the discharge occurs at or above ½ the bias voltage. In accordance, the charge released by the current source was set to a magnitude corresponding to ½ the stored charge on the cell. The duration of the current pulse is determined by avalanche formation and electron transport time across the discharge gap and set to a few ps. There are a number of important results from the SPICE simulation:

1) The potential across an activated pixel drops to ½ the supply value in less than 10 ps.
2) No other cell in the chain, and in particular no neighboring cells, experience any drop of the high voltage.
3) Signal formation time is just a few ps with a magnitude exceeding 1 V.
4) Fall time constant is ~ 240 ps. Therefore the duration for 99.9 % full cell recovery to the baseline is < 2 ns.
5) The integrated energy dissipation for a single pixel discharge is about 56 pJ, modestly larger than obtained for the back-of-envelope calculation. This sets the power consumption at < 2 μW/cm$^2$.

## V. DESIGN OF PPS TEST CHAMBER

The test chamber is a vacuum vessel with ports for gas supply, gas exhaust, and electrode feed-through. A simplified sketch of this test chamber is shown in Figure 10. A 25 cm$^2$ platform, translatable along the vertical axis towards the drift electrode window, serves as the stage upon which PPS test cells are to be mounted. The PPS test cells will be fabricated using multichip module ceramic technology on low-cost 4.9" x 5.4" alumina substrates (i.e. 99.5% $Al_2O_3$, and 0.5 mm thick). On the top surface will be thin-film electrodes of 4 different pitches. Electrodes connect to the "bottom" plane of the substrate which uses thick film technology with controlled resistivity. The cell discharge resistances are printed and laser trimmed in the bottom plane film.

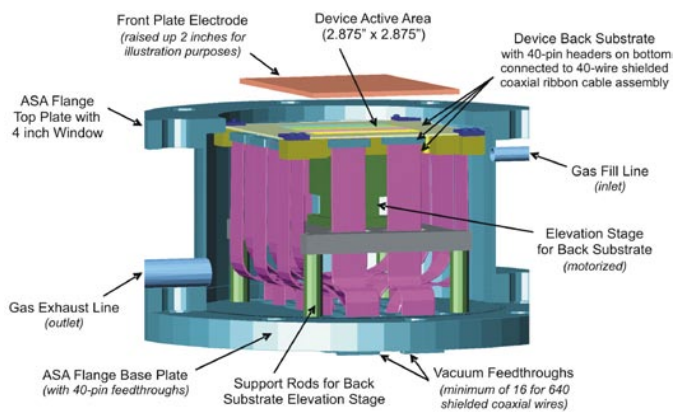

**Figure 10:** Design of PPS test vessel under construction, showing front plate (drift electrode), sectored back plate active pixel detection area, shielded coaxial ribbon signal wiring, signal feedthroughs, gas ports, motorized back plate adjustable gas gap mechanism, aperture window and chamber structure.

The laboratory experimental program includes testing a series of relatively small PPS devices. Each test device will have an active area of ~ 53 cm$^2$, and is being sectored into four differently designed discharge regions of about 13 cm$^2$ each. For the first fabricated devices, the respective pixel pitches will be on the order of a PDP, but will allow a safe initial test and evaluation of the basic device discharge performance and electrode fabrication uniformity. The pixel pitches are 1.02 mm and 0.34 mm (i.e. 25 and 75 pixels/inch), with each area having two different internal sense electrode and discharge electrode widths and discharge gaps, and two different Z-electrode dimensions and pitches. By evaluating two different pixel resolutions, each with two different internal structures and two different Z-electrode configurations on the same substrate, the measured response of these eight different regions can be used to calibrate our simulations, and reasonably estimate the performance of the planned 100 µm pixel resolution devices.

The test chamber design allows investigation, in addition to micron step adjustments of the PPS back substrate to drift electrode distance (i.e. from ~ 50 microns to 1 cm), of different electrode structures and materials, substrates (e.g. glasses, ceramics, semiconductors and metal foils), and gas mixtures. The open test-cell design approach thus provides maximum experimental flexibility to change virtually every PPS device parameter. The test chamber will have 16 custom designed, high-voltage vacuum electrical feedthroughs, each designed to couple to two 20-wire shielded coaxial ribbon cables to accommodate up to 640 PPS electrodes in the planned 100 µm pixel resolution devices. The test chamber will incorporate a vacuum pumping station to facilitate quickly replacing one gas mixture with another.

A gas mixing station will allow up to four different gases at any one time to be introduced into the test chamber, and can be backfilled to any reasonable pressure from vacuum up to four atmospheres of *positive* pressure. It is anticipated that for each gas mixture, approximately a half-dozen device operating pressures will be tested before evacuating and changing to another gas mixture. Candidate initial Penning gases to be evaluated will be based on various two and three component mixtures of the following gases: Ar, Ne, Kr, He, Xe, $N_2$, $CO_2$, $CH_4$ and $CF_4$.

This work was supported in part by the United States Government under DOE-SBIR grant: DE-FG02-07ER84749.